\documentclass[british,a4paper]{amsart}
\usepackage{lmodern}
\usepackage[T1]{fontenc}

\usepackage{babel}
\usepackage[useregional]{datetime2}
\DTMdefzonemap{05}{30}{IST}

\usepackage{amsmath,amssymb,amsthm}
\usepackage[dvipsnames]{xcolor}
\definecolor{brickred}{RGB}{203,65,84}
\usepackage[colorlinks, linkcolor=MidnightBlue, citecolor=MidnightBlue, urlcolor=brickred, linktocpage]{hyperref}
\hypersetup{
	pdfstartview={FitH},    
	pdftitle={Replicas for Random Matrices},    
	pdfauthor={Madhusudhan Raman},     
}

\usepackage{amsaddr}
\usepackage{parskip}
\usepackage{tikz-cd}
\numberwithin{equation}{section}
\usepackage[vcentermath]{youngtab}

\newcommand{\p}{\partial}
\newcommand{\ii}{\mathrm{i}}
\newcommand{\dd}{\text{d}}

\newcommand{\bracket}[1]{\left\langle #1 \right\rangle}

\usepackage{cleveref}
\usepackage[shortalphabetic]{amsrefs}

\title{Replicas for Random Matrices}
\author{Madhusudhan Raman}
\address{International Centre for Theoretical Sciences\\Tata Institute of Fundamental Research\\Shivakote, Hesaraghatta Hobli, Bengaluru 560 089, India}
\curraddr{Instituto de Física Teórica, UNESP - Universidade Estadual Paulista, R. Dr. Bento Teobaldo Ferraz 271, São Paulo 01140-070, Brazil}
\email{madhusudhan dot raman at unesp dot br}
\date{\DTMnow}

\begin{document}

\begin{abstract}
	We discuss the use of the replica ansatz in computing free energies in random matrix theory, and confirm a conjectured condition on analytic continuation in the replica index at large-$ N $.
\end{abstract}

\maketitle
\tableofcontents

\section{Introduction}
In this short note we show that at low temperatures, the free energy of a Gaussian random matrix theory is specified by the left edge of the Wigner semicircle distribution. We do this using the replica ansatz. 

In this section we will situate this result in the context of contemporary research into the connections between random matrix theory and two-dimensional gravity.
 
\subsection*{Gravitational Path Integrals}
Recent work \cite{Saad:2019lba} has established a striking correspondence between the two-dimensional dilaton gravity \cite{Jackiw:1984je,Teitelboim:1983ux} on one hand, and a double-scaled matrix model with a Schwarzian density of states on the other. This demonstration invites us to think of Jackiw-Teitelboim (JT) gravity as holographically dual to an \emph{ensemble} of random Hamiltonians.

Concurrently, the significance of Euclidean replica wormholes to the black hole information paradox --- specifically, the observation that accounting for wormhole saddles between disconnected boundaries allows one to derive a unitary Page curve --- was highlighted by \cite{Penington:2019kki} and \cite{Almheiri:2019qdq}. This was similarly striking, as the inclusion of wormhole saddles would \emph{prima facie} preclude the factorisability of the gravitational path integral on disconnected boundaries. Once again, as discussed for example by \cite{Harlow:2018tqv}, we find that the interpretation of a gravitational path integral as computing an \emph{ensemble} average is an attractive, if curious, possibility.

These developments were further clarified to us in a beautiful paper \cite{Engelhardt:2020qpv}, which set out to understand how the gravitational path integral is influenced by Euclidean wormhole solutions. Their argument is worth recalling. Consider a gravitational path integral $ P(B) $ anchored to a boundary $ B $:
\begin{equation}\label{key}
	P(B) = \int_{\p M = B} \mathrm{D}g \, e^{-S} \ ,
\end{equation}
where we integrate over metrics $ g $ on $ M $ such that $ \p M = B $. For the case of $ m $ disconnected boundaries $ B^{m} = B \cup \cdots \cup B $, we similarly define $ P(B^{m}) $. Now, if Euclidean wormholes contribute to this partition function, then we must have
\begin{equation}\label{key}
	P(B^{m}) \neq P(B)^{m} \ .
\end{equation}
This is what one means when one says that the gravitational path integral fails to factorise. This failure is naturally accommodated by the proposal that the gravitational path integral computes an average over some ensemble of theories, since it is well known that the average of products is not the same as the product of averages. 

This failure to factorise is seen, explicitly, in the difference between ``quenched'' $ (F_{\mathrm{q}}) $ and ``annealed'' $ (F_{\mathrm{a}}) $ free energies of these theories:
\begin{equation}\label{key}
	\begin{aligned}
		F_{\mathrm{q}} &\sim \bracket{\log Z(\beta)} \ , \\
		F_{\mathrm{a}} &\sim \log \bracket{Z(\beta)} \ ,
	\end{aligned}
\end{equation}
where $ \bracket{\cdots} $ denotes an ensemble average. Working with $ \widehat{\mathrm{CGHS}} $ and JT gravities, the authors of \cite{Engelhardt:2020qpv} observe that while at high temperatures the quenched and annealed free energies agree, at low temperatures they begin to differ considerably due to the increasingly dominant contributions of (wormhole) connected correlators. This is also the regime where the annealed free energy exhibits pathological behaviour: it is not monotonically decreasing, thereby implying a negative thermodynamic entropy.

Interestingly, even after the contributions of replica wormholes are taken into account at low temperatures, and although this inclusion goes some way towards easing the tension signalled by non-monotonicity, the (quenched) free energy continues to behave pathologically.

\subsection*{Toy Models For Averaging}
In light of these developments, attempts have been made to understand averaging \emph{in vitro}, so to speak, in the relatively simple context of random matrix theory. Notable among these and germane to our theme is \cite{Okuyama:2020mhl}, which studies ``small'' (rather, finite-$ N $) random matrices using a combination of analytic and numerical methods, and finds that the quenched free energy exhibits no signs of pathological behaviour, viz.~it decreases monotonically, ensuring a positive entropy at all temperatures. This finite-$ N $ eigenvalue model, the workhorse of \cite{Okuyama:2020mhl}, suggests that at low temperatures, the free energy is equal to the smallest eigenvalue of the matrix ensemble.

The questions we explore in this short note are: does this remain true at large-$ N $? And can the replica ansatz be used to demonstrate the same? After reviewing the necessary background in Section \ref{sec:Review}, we answer both these questions in the affirmative in Section \ref{sec:Results}.

\subsection*{Random Energy Model and Rectangular Diagrams}
We are motivated to pursue this question on the strength of a similarity between this problem and an exactly solvable model of disordered systems called the random energy model \cite{Derrida:1981zz}. Correlators of $ n $-point functions of the partition function in both the random energy model and random matrix theory, as we will describe in more detail in subsequent sections, can be written as a sum of $ p(n) $ terms, where $ p(n) $ is defined as
\begin{equation}\label{key}
	\prod_{k=1}^{\infty} \frac{1}{1-x^{k}} = \sum_{n=0}^{\infty} p(n) x^{n} \ ,
\end{equation}
and counts the number of partitions of $ n $. Each of these $ p(n) $ terms is associated canonically to a Young diagram corresponding to a partition of $ n $. In the random energy model, at high temperatures the totally disconnected partition $ n =1+\cdots + 1 $ dominates, and the replica ansatz for computing the quenched free energy
\begin{equation}\label{key}
	\bracket{\log Z} = \lim_{n \rightarrow 0} \frac{\bracket{Z^n}-1}{n} \ ,
\end{equation}
works swimmingly, since $ \bracket{Z^{n}} \simeq \bracket{Z}^{n} $. At low temperatures, where connected correlators begin to compete for dominance, the replica trick naively fails to return sensible results.

It turns out that the way to extract physically sensible results using the replica ansatz is to assume that some rectangular Young diagram is dominant at low temperatures, and solve for its width by requiring that the corresponding free energy is extremised. In the language of Parisi's replica symmetry breaking (RSB) scheme (see \cite{Mezard:1987} for a thorough discussion and a useful collection of reprints) this is essentially the ``$ 1 $-step RSB.'' We will adopt this strategy when studying the matrix model as well.

Our new results are: (i) a much simpler closed-form expression for the generating function of $ n $-point connected correlators in \cref{eq:GeneratingFunctionConnectedCorrelators}, and (ii) a proof that this $ n $-point functions at large-$ N $ and low temperatures satisfies the following condition on analytic continuation:
\begin{equation}\label{key}
	\lim _{\beta \rightarrow \infty}\left\langle Z(\beta)^{n}\right\rangle=e^{-n \beta E_{0}} \ .
\end{equation}
In the above formula, $ E_{0} = -2 $, the left edge of the Wigner semicircle distribution. This condition was first proposed by \cite{Okuyama:2020mhl} and we prove this using the replica ansatz.

\subsection*{Acknowledgments}
The author is grateful to Sujay Ashok, Diptarka Das, Shouvik Datta, Oliver Janssen, and Dileep Jatkar for discussions, and to Rukmini Dey and Suvrat Raju for arranging a short-term visiting position at the International Centre for Theoretical Sciences.

\section{Exact Results}
\label{sec:Review}
As we discussed in the previous section, it is our goal in this paper to explore the use of the replica ansatz in computing free energies in random matrix theory. Concretely, we consider a system governed by a random Hamiltonian $ H $ drawn from a Gaussian unitary ensemble (GUE). Each instance of this ensemble is associated to a partition function
\begin{equation}\label{key}
	Z(\beta) = \operatorname{Tr} e^{-\beta H} \ ,
\end{equation}
where $ H $ is a Hermitian $ N \times N $ matrix and $ \beta = T^{-1} $ is the inverse temperature. Quantities of interest in this theory are averaged over the GUE like so:
\begin{equation}\label{key}
	\bracket{f(H)} = \frac{1}{\mathsf{Z}} \int \dd H \, e^{-\frac{N}{2} \operatorname{Tr} H^{2}} f(H) \ ,
\end{equation}
where $ \mathsf{Z} $ is defined such that $ \bracket{1} = 1 $. In particular, the correlation function we are interested in is the ``quenched'' free energy $ \bracket{\log Z(\beta)} $, which according to the replica ansatz is given by
\begin{equation}\label{key}
	\langle\log Z(\beta)\rangle=\lim _{n \rightarrow 0} \frac{\left\langle Z(\beta)^{n}\right\rangle-1}{n} \ .
\end{equation}
For this reason, the objects of central interest to us in this note will be $ \bracket{Z(\beta)^{n}} $, which we will occasionally call $ n $-point functions without qualification. 

Happily, the correlation functions $ \bracket{Z(\beta)^{n}} $ in random matrix theory are well-known from the study of $ \tfrac{1}{2} $-BPS Wilson loops in four-dimensional $ \mathcal{N} = 4 $ super Yang-Mills theory, as was highlighted in \cite{delCampo:2017bzr,Okuyama:2018yep}. Formally, the results of these computations are related by the replacement
\begin{equation}\label{key}
	\sqrt{\lambda} \leftrightarrow 2\beta \ ,
\end{equation}
where $ \lambda $ is the 't Hooft coupling of the four-dimensional gauge theory. In light of this correspondence, we will frequently use results from the gauge theory and the matrix model interchangeably. 

It will be useful for us to observe that for any integer $ n $, the correlation functions $ \bracket{Z(\beta)^{n}} $ may be decomposed into sums of connected components --- identified as $ \bracket{\,\cdot\,}_{c} $ --- and each contribution to this decomposition is labelled by a partition of the integer $ n $, which is canonically associated to a Young diagram
\begin{equation}\label{key}
	Y = \left[ 1^{k_1} \cdots j^{k_j} \cdots n^{k_n} \right] \ , 
\end{equation}
which, for all $ j \in \{1,\cdots,n\} $, has $ k_j $ rows of $ j $ boxes. Of course, we will require that $ |Y| $, the total number of boxes, is $ n $:
\begin{equation}\label{key}
	|Y| = \sum_{j} j k_{j} = n \ .
\end{equation}
With this in place, the correlation function is decomposed as
\begin{equation}\label{key}
	\left\langle Z(\beta)^{n}\right\rangle=\sum_{|Y|=n} Z_{Y}(\beta) \ ,
\end{equation}
where each contribution labelled by a partition of $ n $ (alternatively, an $ n $-box Young diagram) is given by
\begin{equation}\label{key}
	Z_{Y}(\beta)=n ! \, \prod_{j=1}^{n} \frac{1}{k_{j} !}\left[\frac{\left\langle Z(\beta)^{j}\right\rangle_{c}}{j !}\right]^{k_{j}} \ .
\end{equation}
The $ n $-point function thus receives contributions from all partitions of $ n $. As the temperature is varied, the various partitions of $ n $ jockey for dominance within the sum, and it is conceivable that there will exist some temperatures at which dominance is traded. A simpler and more explicit version of this phenomenon is presented in \cite{Engelhardt:2020qpv}, where in their discussion of the $ \widehat{\mathrm{CGHS}} $ model, they show that at lower temperatures connected correlators dominate the partition sum. We will briefly discuss this theory in the final section.

The $ 1 $-point function $ \bracket{ Z(\beta) } $ in random matrix theory was computed exactly at finite $ N $ \cite{Drukker:2000rr} and it is given in terms of associated Laguerre polynomials as
\begin{equation}\label{eq:FiniteN1PointFunction}
	\langle Z(\beta)\rangle=e^{\frac{\beta^{2}}{2 N}} L_{N-1}^{1}\left(-\frac{\beta^{2}}{N}\right) \ .
\end{equation}
For future convenience, let us define
\begin{equation}\label{eq:AZBeta}
	A_{k} = \bracket{Z(k \beta)} \ .
\end{equation}
Higher-point correlators of the partition function can also be computed in terms of these $ A_k $, as was demonstrated in \cite{Okuyama:2018aij} through the following contour integral formula:
\begin{equation}\label{eq:OkuyamaMasterContourIntegralFormula}
	\left\langle Z(\beta)^{n}\right\rangle_{c}=\oint \prod_{i=1}^{n} \frac{\dd z_{i}}{2 \pi \ii \, z_{i}^{2}} \operatorname{Tr} \log \left[\sum_{m=0}^{n} \sum_{i_{1}<\cdots<i_{m}} z_{i_{1}} \cdots z_{i_{m}} A_{m} \right] \ .
\end{equation}

We now propose a much simpler formula that captures the same content as the above equation:
\begin{equation}\label{eq:GeneratingFunctionConnectedCorrelators}
	\left\langle e^{t Z(\beta)}\right\rangle_{c}=1+\log \left[1+\sum_{k=1}^{\infty} A_{k} \frac{t^{k}}{k !}\right] \ ,
\end{equation}
Observe that the l.h.s.~of \cref{eq:GeneratingFunctionConnectedCorrelators} is in fact the generating function of connected correlators of the partition function. The above expression allows us to relate connected higher-point correlators of the partition function to the $ A_{k} $, in addition to sidestepping the difficulty of performing a multi-dimensional contour integral. In particular, by expanding each side in a Taylor series around $ t = 0 $, we recover the same content as \cref{eq:OkuyamaMasterContourIntegralFormula}. For example, for low $ k $ we find by simple Taylor expansions
\begin{equation*}\label{key}
	\begin{aligned}
		\bracket{Z(\beta)^3}_{c} &= A_{3} - 3 A_{1} A_{2} + 2 A_{1}^3 \ , \\
		\bracket{Z(\beta)^4}_{c} &= A_{4} - 4 A_{1} A_{3} - 3 A_{2}^2 + 12 A_{1}^2 A_{2} - 6 A_{1}^{4} \ , \\
		\bracket{Z(\beta)^5}_{c} &= A_{5} - 5 A_{1} A_{4} - 10 A_{2} A_{3} + 20 A_{1}^{2} A_{3} + 30 A_{1} A_{2}^{2} - 60 A_{1}^3 A_{2} + 24 A_{1}^{5} \ ,
	\end{aligned}
\end{equation*}
and it is easily verified that these expressions match the results of the contour integral representation.

That \cref{eq:GeneratingFunctionConnectedCorrelators} is correct is easy to see once we recognise that it is simply an ``inversion'' of the standard relation between a correlation function and its connected part:
\begin{equation}\label{eq:MomentGeneratingFunction}
	\left\langle e^{t Z(\beta)}\right\rangle=\sum_{n=0}^{\infty} \frac{t^{n}}{n !}\left\langle Z(\beta)^{n}\right\rangle=\exp \left[\sum_{j=1}^{\infty} \frac{t^{j}}{j !}\left\langle Z(\beta)^{j}\right\rangle_{c}\right] \ .
\end{equation}
The above relation tells us that the correlation function $ \bracket{Z(\beta)^n} $ is given by the $ n $-th Bell polynomial $ B_{n}(x_{1},\cdots,x_{n}) $ in the connected correlation functions
\begin{equation}\label{key}
	x_{k} = \bracket{Z(\beta)^k}_{c} \ .
\end{equation}
Essentially, where the $ \bracket{Z(\beta)^{n}} $ are like moments of a distribution, the connected correlation functions $ \bracket{Z(\beta)^n}_{c} $ are like cumulants of a distribution.\footnote{See \cite{McCullagh:2009} for more details.}

It will be useful for us to recall also the large-$ N $ limits of \cref{eq:FiniteN1PointFunction}. It was first observed in \cite{Erickson:2000af} that 
\begin{equation}\label{eq:ZBetaLargeN}
	\langle Z(\beta)\rangle=\frac{N}{\beta} I_{1}(2 \beta) \ ,
\end{equation}
where $ I_{\nu}(z) $ is a modified Bessel function. Since we have succeeded in expressing all connected $ n $-point correlators in terms of the $ A_{k} $ using \cref{eq:AZBeta} and \cref{eq:GeneratingFunctionConnectedCorrelators}, this result is sufficient for our purposes.

\section{Replica Ansatz}
\label{sec:Results}
Our interest is in computing the quenched free energy using the replica ansatz:
\begin{equation}\label{eq:ReplicaAnsatz}
	\langle\log Z(\beta)\rangle=\lim _{n \rightarrow 0} \frac{\left\langle Z(\beta)^{n}\right\rangle-1}{n} \ .
\end{equation}
The trouble with implementing the replica ansatz, as we have discussed, arises at low temperatures. Before discussing this, we quickly recall some results at high temperature.

\subsection*{High Temperatures}
Let us first consider the behaviour of $ \bracket{Z(\beta)^{n}} $ at high temperatures. In this regime, the dominant contribution is the totally disconnected correlator corresponding to the partition $ Y = \left[1^{n}\right] $. Thus, at high temperatures we have
\begin{equation}\label{key}
	\bracket{Z(\beta)^{n}} \simeq \bracket{Z(\beta)}^n \ ,
\end{equation}
and the replica ansatz \cref{eq:ReplicaAnsatz} finds that
\begin{equation}\label{key}
	\bracket{\log Z(\beta)} \simeq \log \bracket{Z(\beta)} \ .
\end{equation}
That is, at high temperatures the quenched and annealed free energies agree and the replica ansatz returns an answer that is physically reasonable; as \cite{Okuyama:2020mhl} notes and as one can observe from \cref{eq:ZBetaLargeN}, in the the limit $ \beta \rightarrow 0 $ we have
\begin{equation}\label{key}
	\lim_{\beta \rightarrow 0} Z(\beta) = N \ ,
\end{equation}
and so the quenched free energy computed using \cref{eq:ReplicaAnsatz} approaches $ \log N $, the maximal entropy of the system.

\subsection*{Low Temperatures}
At low temperatures, however, the application of the replica ansatz frustrated by well-known ambiguities involving analytic continuation. For example, it was shown through the numerical studies of \cite{Okuyama:2019xvg} that at low temperatures the dominant contribution to the $ n $-point function comes from the fully connected correlator labelled by $ Y = \left[n^{1}\right] $, and this in turn implies the following simplification:
\begin{equation}\label{key}
	\bracket{Z(\beta)^{n}} \simeq A_{n} = \bracket{Z(n\beta)} \ .
\end{equation}
Unfortunately, when the above simplification is plugged into the replica ansatz in \cref{eq:ReplicaAnsatz}, the limit is ill-defined and in this way the replica ansatz appears to fail. The temperature at which quenched and annealed quantities begin to differ --- a signature of what is referred to as replica symmetry breaking --- is typically denoted $ T_{\mathrm{RSB}} $.

It will be important in subsequent sections to note that the demonstration of connected correlator dominance at low temperatures was carried out for integer $ n $. This is interesting in its own right, but for the purposes of the replica ansatz is inconclusive since the replica ansatz necessitates a departure from integer $ n $. In other words: the dominance of $ Y = \left[n^{1}\right] $ for integer $ n $ in no way guarantees its dominance as $ n $ is analytically continued away from the integers.

The goal of the subsequent sections will be to argue that the replica ansatz can in fact return physically reasonable answers at low temperatures. For this, we must first chart the landscape of contributions to the free energy.

\subsection*{Parametrising The Landscape}
In \cref{eq:GeneratingFunctionConnectedCorrelators}, we have seen how to express connected correlation functions in terms of sums of products of the functions $ A_{k} $. Schematically, each contribution to an $ n $-point function is, schematically, a product of $ p $ factors of the form
\begin{equation}\label{key}
	A_{k_1} \cdots A_{k_p} \quad \text{such that} \quad \sum_{i=1}^{p} k_{i} = n \ .
\end{equation}
The replica ansatz requires that we analytically continue from $ n \in \mathbb{Z}_{+} $ to $ n \in \mathbb{R}_{\geq 0} $; crucially, we will retain the above constraint on the $ k_{i} $, i.e.~that they sum to $ n $.

An important observation of \cite{Okuyama:2019xvg} is that near the the temperature at which replica symmetry breaking occurs, all partitions of $ n $ contribute in roughly equal measure to any computation of the free energy. There is, thus, at $ T_{\mathrm{RSB}} $, a landscape of free energies which contribute in roughly equal magnitude to the partition sum, and there is no clear dominance of one partition over another.

This picture of the landscape of free energies is productive because at fixed integer $ n $, we can explicitly see an exchange of dominance from totally connected to totally disconnected correlation functions as the temperature is ramped up. At fixed $ \beta $, however, varying $ p $ essentially interpolates between different partitions.

In light of this, we suppose that within the replica ansatz, it is \emph{a priori} unclear which partition dominates at low temperatures. (This is to be contrasted with the case of integer $ n $, where we know that the partition $ \left[n^1\right] $ dominates.) We should then attempt to parametrise the landscape of contributions to the free energy and extremise it with respect to these parameters.

The simplest such parametrisation is to suppose that a ``rectangular'' partition
\begin{equation}\label{key}
	\underbrace{\left(p,\,\cdots,p\right)}_{\frac{n}{p}\,\mathrm{times}}
\end{equation}
is dominant. This proposal takes a leaf out of earlier work on the random energy model \cite{Derrida:1981zz} and was more recently suggested by \cite{Okuyama:2020mhl}. Observe that $ p = 1 $ corresponds to the totally disconnected correlator (dominant at high temperatures), $ p = n $ corresponds to the totally connected correlator (dominant at low temperatures when $ n $ is an integer), and intermediate values of $ p $ correspond to other allowed terms. It is clear that $ p $ parametrises the shape of the Young diagram, and such a term contributes the following to the $ n $-point function:
\begin{equation}\label{key}
	\left( A_{p} \right)^{n/p} \ .
\end{equation}
For some $ p = p_{\star} $ this quantity will be extremal --- this contribution, when inserted into the replica ansatz, ought to correspond to the correct free energy.

\subsection*{Low Temperatures Revisited}
Let us try and compute the quenched free energy at large $\beta$ (low temperatures) using the technique we have just discussed. We start with
\begin{equation} 
\label{eq:ArbitraryP}
	\left( A_{p} \right)^{n/p} \simeq \left[ \frac{N}{(p\beta)^{3/2}} e^{2p\beta}\right]^{n/p} \ ,
\end{equation} 
where we have ignored numerical factors that appear in the asymptotic expansion for the modified Bessel functions since they will not matter. Extremising this with respect to $p$, we find that 
\begin{equation} 
p_{\star} = e\left( \frac{N^{2/3}}{\beta} \right) \ ,
\end{equation} 
which when we plug back into \cref{eq:ArbitraryP} gives us
\begin{equation} 
\left( A_{p_{\star}} \right)^{n/p_{\star}} = \exp\left\{ n\beta \left( 2- \frac{3}{2eN^{2/3}} \right) \right\} \ .
\end{equation} 
Now, using the standard thermodynamic prescription
\begin{equation} 
\begin{aligned}
	F &= -\beta^{-1} \langle \log Z \rangle \ , \\
	&= -2+\frac{3}{2eN^{2/3}} \ , \\
	&\simeq -2 \ ,
\end{aligned}
\end{equation} 
where in the last line we have dropped terms that are negligible in the limit $N\rightarrow \infty$.

This is in fact consistent with expectations first elaborated in \cite{Okuyama:2020mhl}: it is natural to expect that for finite $ N $ and $ \beta\rightarrow\infty $ (low temperatures) only the smallest eigenvalue of $ H $ will contribute to
\begin{equation}\label{key}
	F = -\lim_{\beta \rightarrow \infty} \beta^{-1}\bracket{\log Z(\beta)} = - \lim_{\beta \rightarrow \infty} \beta^{-1} \bracket{\log \sum_{i=1}^{N}e^{-\beta E_{i}}} = E^{(N)}_{\mathrm{min.}} \ ,
\end{equation}
and further, that at large-$ N $, $ E^{(N)}_{\mathrm{min.}} $ approaches the left edge of the Wigner semi-circle distribution, which in our matrix integral conventions is $ E^{(\infty)}_{\mathrm{min.}} = E_{0} = -2 $, so
\begin{equation} 
\lim_{\beta \rightarrow \infty} F = E_{0} \ .
\end{equation} 

This computation may be interpreted as confirmation at large-$N$ of a condition on analytic continuation in the replica index $ n $ 
\begin{equation}\label{key}
	\lim _{\beta \rightarrow \infty}\left\langle Z(\beta)^{n}\right\rangle=e^{-n \beta E_{0}} \ ,
\end{equation}
originally proposed in \cite{Okuyama:2020mhl}. While the original condition was explicitly tested at finite $N$ using a combination of analytic and numerical methods, the above can be seen as a large-$N$ version of the same demonstration that successfully employs the replica ansatz. 

\section{Discussion}
Before we conclude, it is interesting to consider the case of $ \widehat{\mathrm{CGHS}} $ gravity and ask if a similar strategy would work there too. We recall some basic facts about this dilaton gravity, and refer the reader to \cite{Godet:2021cdl} for more details. 

The original CGHS model is a dilaton gravity theory in two dimensions \cite{Callan:1992rs}. Solutions to this theory are flat, and have a fixed temperature determined by the cosmological constant and the boundary value of the dilaton. To get solutions with different values of temperatures, following \cite{Afshar:2019axx} we use the action
\begin{equation}\label{key}
	I_{\widehat{\mathrm{CGHS}}}=-\frac{\kappa}{2} \int \dd^{2} x \sqrt{g} \, \left(\Phi R+2 \Lambda-2 \Lambda \varepsilon^{\mu \nu} \partial_{\mu} A_{\nu}\right) \ ,
\end{equation}
and note that the above action reproduces the dynamics of the original CGHS model, except this time $ \Lambda $ can take any constant value, thereby allowing for solutions for varying temperatures. A judicious choice of boundary conditions selects geometries with asymptotic boundaries that are thermal circles, and since dilaton path integral constrains these geometries to be locally flat, they can only be discs or cylinders. The full gravitational path integral (with an appropriate boundary action and a topological term that controls the topological expansion) on a manifold with $ n $ boundaries can be computed just as in the case of JT gravity.

The connected correlators in this theory were computed in \cite{Godet:2020xpk} and are given by
\begin{equation}\label{eq:CGHSGaussian}
	\bracket{Z(\beta)}=\frac{2 \pi}{\beta^{2}} \quad,\quad \bracket{Z(\beta)^2}_{c}=\frac{2 \pi^{2}}{\beta} \quad,\quad \bracket{Z(\beta)^{n\geq3}}_{c} = 0 \ ,
\end{equation}
since there are no flat connected surfaces with $ n \geq 3 $ connected boundaries. That is, $ Z(\beta) $ can be thought of as a Gaussian-distributed random variable with non-zero mean and variance, both specified by the temperature as above. For such a theory, the quantity $ F_{\mathrm{q}} $ is ill-defined, since $ Z(\beta) $ takes values on the entire real line. For this reason, we \emph{define} the quenched free energy of $ \widehat{\mathrm{CGHS}} $ gravity, following \cite{Godet:2020xpk,Engelhardt:2020qpv}, to be
\begin{equation}\label{eq:FreeEnergyCGHS}
	F_{\mathrm{q}} = -\beta^{-1} \bracket{\log |Z(\beta)|}  \ .
\end{equation}

That $ Z(\beta) $ may be thought of as a Gaussian-distributed random variable is interesting in light of the wormhole network picture of \cite{Okuyama:2019xvg}, where all $ n $-point functions are decomposed in terms of Young diagrams / partitions of $ n $, each corresponding to a product of connected parts. The constraint \cref{eq:CGHSGaussian} tells us that $ (n\geq 3) $-point correlators vanish identically, which means only \emph{restricted} partitions are permitted. For example, while the (unrestricted) partitions of $ n = 4 $ are
\begin{equation*}\label{key}
	\left\{ \quad \tiny\yng(1,1,1,1) \quad,\quad \tiny\yng(2,1,1) \quad,\quad \tiny\yng(2,2) \quad,\quad \tiny\yng(3,1) \quad,\quad \tiny\yng(4) \quad \right\} \ ,
\end{equation*}
we see that the set of (restricted) partitions that contribute to the $ 4 $-point multi-boundary correlators in $ \widehat{\mathrm{CGHS}} $ gravity are
\begin{equation*}\label{key}
	\left\{ \quad \tiny\yng(1,1,1,1) \quad,\quad \tiny\yng(2,1,1) \quad,\quad \tiny\yng(2,2) \quad \right\} \ .
\end{equation*}
In general, for any $ n $, Young diagrams with rows having more than two columns are disallowed. We conclude from this that an $ n $-point function can be written as
\begin{equation}\label{key}
\bracket{Z(\beta)^{n}} = \bracket{Z(\beta)}^{n} \sum_{k=0}^{n} T_{n,k} \, \frac{\bracket{Z(\beta)^{2}}_{c}}{\bracket{Z(\beta)}^{2}} \ ,	
\end{equation}
where
\begin{equation}\label{key}
	T_{n,k} = \frac{n!}{k!\, (n-2k)!\,2^{k}} \ ,
\end{equation}
are Bessel numbers, which tell us how many ways there are to partition $ n $ objects into $ k $ sets, each of size one or two.\footnote{For more details, see \cite{Cheon:2013}.}

This brings us to an important point: the restricted nature of the partitions in $ \widehat{\mathrm{CGHS}} $ gravity means our method of implementing the replica ansatz, which involved a variational problem that solved for the width of dominant rectangular Young diagrams, is unlikely to find purchase in this context. 

Although our work has dealt solely with the replica ansatz, we would be remiss if we did not mention a new, exact relation quantifying the difference between quenched and annealed free energies first reported in \cite{Okuyama:2021pkf} and explicated elegantly in \cite{Johnson:2021rsh}. This relation, which essentially derives from an integral representation of the natural logarithm:
\begin{equation}\label{eq:NaturalLogarithm}
	\log x=\int_{0}^{\infty} \frac{e^{-u}-e^{-u x}}{u} \mathrm{~d} u \ ,
\end{equation}
reads
\begin{equation}\label{eq:OkuyamaFormula}
	\langle\log Z\rangle=\log \langle Z\rangle-\int_{0}^{\infty} \frac{\mathrm{d} u}{u}\left[\left\langle e^{-Z u}\right\rangle-e^{-\bracket{Z} u}\right] \ .
\end{equation} 
It has been suggested that the term quantifying the difference between quenched and annealed free energies --- which, as we will see, cleverly encodes the contributions of connected correlators --- corresponds to a superposition of spacetime D-branes \cite{Marolf:2020xie}. Further, subsequent work \cite{Alishahiha:2020jko,Janssen:2021mek,Johnson:2021zuo} explicitly finds that the quenched free energy of JT gravity and its deformations is well-behaved, i.e.~is monotonically decreasing.

We can, for example, still use \cref{eq:OkuyamaFormula} to compute the quenched free energy of $ \widehat{\mathrm{CGHS}} $ gravity as defined above exactly. This computation is simplified by the fact that only connected correlators factor into the term quantifying the difference between quenched and annealed free energies, and as we have seen, almost all of them vanish. A simple calculation, which requires analytic continuation from $ \beta < 0 $ followed by picking out the real part, fully analogous to the prescription implied by \cref{eq:FreeEnergyCGHS}, finds
\begin{equation}\label{key}
	F_{\mathrm{q}} = F_{\mathrm{a}} + \frac{T}{2} \left[ \gamma + \log 4T^{3} -2 T^{3} \times {}_{2}F_{2} \left(1,1;\tfrac{3}{2},2;-T^{3}\right) \right] \ .
\end{equation}
This confirms a result of \cite{Godet:2021cdl}.

\bibliographystyle{amsplain}
\bibliography{Refs}
\end{document}